# Explainable unsupervised multi-modal image registration using deep networks

Chengjia Wang [1, 2], Giorgos Papanastasiou [2, 3]

1. School of Mathematical and Computer Sciences, Heriot-Watt University, Edinburgh EH14 4AS, UK
2. Edinburgh Imaging Facility-QMRI, University of Edinburgh, Edinburgh EH16 4TJ, UK
3. School of Computer Science and Electronic Engineering, University of Essex, Colchester CO4 3SQ, UK

**ABSTRACT**

Clinical decision making from magnetic resonance imaging (MRI) combines complementary information from multiple MRI sequences (defined as "modalities"). MRI image registration aims to geometrically "pair" diagnoses from different modalities, time points and slices. Both intra- and inter-modality MRI registration are essential components in clinical MRI settings. Further, an MRI image processing pipeline that can address both afine and non-rigid registration is critical, as both types of deformations may be occuring in real MRI data scenarios. Unlike image classification, explainability is not commonly addressed in image registration deep learning (DL) methods, as it is challenging to interpet model-data behaviours against transformation fields. To properly address this, we incorporate Grad-CAM-based explainability frameworks in each major component of our unsupervised multi-modal and multi-organ image registration DL methodology. We previously demonstrated that we were able to reach superior performance (against the current standard Syn method). In this work, we show that our DL model becomes fully explainable, setting the framework to generalise our approach on further medical imaging data.

## 1. INTRODUCTION

Magnetic resonance imaging (MRI)-based clinical decision making benefits from combining complementary anatomical and functional information, from multiple MRI sequences (defined here as "modalities"). Multiple visual assessments and imaging biomarkers can be derived from different MR modalities, to synergically form a single diagnosis. Image registration is an important MRI analysis process, as it is important to geometrically "pair" images from different modalities, time points and slices. Thus, intra- and inter-modality image registration are essential components in clinical MR image analysis [1, 2].

Despite numerous deep learning (DL) methods were developed for medical image analysis [2], DL-based image registration tasks have been relatively less investigated [3, 4]. Although challenging, unsupervised DL-based image registration methods started gaining increasing popularity, as they focus to overcome the need of manually drawn annotations (needed as ground truths) in supervised learning techniques [3, 4], which are laborious and time-consuming [5].

However, most previous unsupervised DL methods can perform either affine or non-rigid registration [3, 4, 6-11], on 3D volumes [6-9] or 2D images [10, 11], whilst mainly focus on intra(single)-modal image registration [3, 4]. To our knowledge, the only versatile unsupervised DL study that has developed a method for both affine and non-rigid registrations is by de Vos et al [12]. In this study, the authors developed two autonomous models to perform affine and non-rigid registration, whilst also required affine transformations before training which is a laborious, time-consuming and computationally expensive task [12].

Furthermore, it is important to emphasise that explainability has been less explored in image registration DL tasks. One of the main reasons is the lack of ground truths and the challenge to unravel model-data behaviours against transformation field adaptations, during image registration tasks. Two previous studies reported gradient-weighted class activation map (Grad-CAM)-based methodologies [13] to perform DL vizualisations through gradient-based localization, in image registration pipelines [14, 15]. Wei et al. involved Grad-CAM to facilitate pre-procedural CT to intra-procedural CT registration, through probe localization and inpainting, for thermal ablation [14]. Song et al. employed Grad-CAM to better understand the function of cross-modal attention blocks in their model, for MRI and Ultrasound image registration [15].

We previously demonstrated a robust DL method (named "FIRE") that was able to dynamically learn anatomical and latent representations, simultaneously across modalities [16]. We had showed that our method efficiently models inter- and intra-modality image registration, through bi-directional cross-domain synthesis and factorised spatial transformations [16]. For the first-time, we focus here on enhancing model explainability by incorporating Grad-CAM pipelines within each of our model components. We name our proposed method as "ExFIRE". By deriving Grad-CAM visualizations, we fully explicitly explain our model performance and set the framework to generalize our method for additional medical imaging data.

In summary, our contributions are as follows: a) We propose to incorporate Grad-CAM methods within each of the components of our previous FIRE model for unsupervised cross-modal image registration, to elucidate unsupervised cross-modal registration mechanisms; b) We unravel how each of the model components contribute to the overall cross-modal registration tasks, while addressing inverse-consistency throughout our method; c) We show the explainability of our method in cross-modal registration, setting the framework for further model generalizations in the broader medical imaging domain.

## 2. RELATED WORK

In this section, we review prior work in the field of DL image registration. We focus on reviewing methods that are related to unsupervised image registration, inverse-consistent DL transformations, and explainability in image registration.

**Unsupervised image registration**

Unsupervised learning approaches gain popularity in image registration, as they can overcome the need of manual ground truth annotations for training. The spatial transformer network (STN, in 2015) development inspired unsupervised learning methods, as it can be engineered within existing DL networks [17]. The STN allows explicit manipulation of images within DL networks. The Voxelmorph model is a 3D medical image registration algorithm that uses STN within a convolutional neural network, learning via cross-correlation loss functions [18]. Krebs et al. developed a probabilistic formulation of unsupervised registration, in which an STN was used within a conditional variational autoencoder [19]. Kim et al. developed CycleMorph, a cycle consistent DL method for deformable image registration (through an STN component), demonstrating robust registration results [20]. Unlike previous techniques [18, 19], the CycleMorph method incorporated inverse consistency directly in the training phase [20], to improve model efficiency and robustness. de Vos et al developed an unsupervised DL study for both affine and non-rigid registrations [12]. However, as mentioned, DL modelling required the incorporation of two autonomous models in the analysis to address both affine and non-rigid registrations, whilst also required affine transformations as part of the modelling process (which added complexity and computational cost).

**Inverse-consistent transformations**

In our previous work, we demonstrated a versatile and efficient DL model that can perform both affine and non-rigid (deformable) registration while intrinsically producing inverse-consistent topology-preserving image registrations [16]. Note that inverse consistency is an important property to consider in registration tasks. In previous studies, diffeomorphic architectures such as the Demons algorithms, are known that can address inverse consistency [21, 22]. Several DL diffeomorphic architectures have recently been devised, such as the Voxelmorph [18], the CycleMorph [20] and the Quicksilver [22]. However, most of these unsupervised diffeomorphic models were examined on intra-modality brain data [18, 22]. Moreover, a recent study showed that the diffeomorphism can not account for local discontinuities in brain images with pathological conditions [23]. In our work, we define the output deformation fields with simple inverse-consistency constraints, devising a model that can learn topology-preserving registration in an expandable and versatile architecture. Further, in the context of explainability, we unravel and carefully explain how inverse consistency occurs within our proposed method.

**Explainability in image registration**

Although Grad-CAM-based explainability methodologies have been incorporated in multiple image classification problems, or to explain medical diagnosis, they have not been investigated in DL image registration [13-15, 24]. Wei et al. devised Grad-CAM to automate probe localization and inpainting, as an intermediate step in the setting of pre-procedural to intra-procedural CT registration [14]. Song et al. incorporated Grad-CAM to understand the function of cross-modal attention blocks in their model, for MRI and Ultrasound image registration [15]. To our knowledge, we are the first to explicitly unravel end-to-end DL model mechanisms through Grad-CAM on image registration. We show that DL explainability is possible even in complex deep networks, as for example in our FIRE model, where parallilised cross-domain image synthesis and factorised spatial transformations are devised to perform bi-directional image registration.

## 3. METHODS

**ExFIRE model**

Details of our FIRE architecture and model processes have been previously described [16]. In brief, the FIRE model consists of a bi-directional cross-domain image synthesis structure, with two factorised spatial transformation components (one per each encoder-decoder channel, Figure 1).

When two images $x^A$ and $x^B$ are processed, the FIRE model learns $\varphi^{A \rightarrow B}$ and $\varphi^{B \rightarrow A}$ transformations, warping $x^A$ and $x^B$ into $x^A \circ \varphi^{A \rightarrow B}$ and $x^B \circ \varphi^{B \rightarrow A}$, through the transformation networks $T^{A \rightarrow B}$ and $T^{B \rightarrow A}$. In parallel, a synthesis encoder extracts modality-invariant latent representations ($G(x^A)$, $G(x^B)$) which then run through two synthesis decoders $F^{A \rightarrow B}$ and $F^{B \rightarrow A}$, mapping the representations extracted by $G(x^A)$ and $G(x^B)$ to the synthesed images $\hat{x}^B$ and $\hat{x}^A$, respectively.

For the ExFIRE model, following the adversarial learning fashion [6], two discriminator networks, Discriminator Net A ($D^A$) and Discriminator Net B ($D^B$), are subsequently trained to differentiate the real from the synthesized images. In the meantime, the other network components are trained to confuse the discriminator networks. Note that the transformed images are considered as real by $D^A$ and $D^B$ (Figure 1).

We generate Grad-CAM-based attention maps for the last convolutional layers of: 1) the encoder network; 2) the two spatial transformation networks; and 3) the two discriminator networks. Mathematically, the Grad-CAM attention maps of the encoder and transformation networks, correlates the predicted transformation fields, to the reference imaging space.

The Grad-CAM-based attention maps of the discriminator networks aim to display similarities or differences in the critical visual clues (Grad-CAM scores) between the real and synthesised images. Essentially, Grad-CAM-based attention maps of the discriminator networks reflect the quality of synthesized images, which in turn further explains the registration procedure performance.

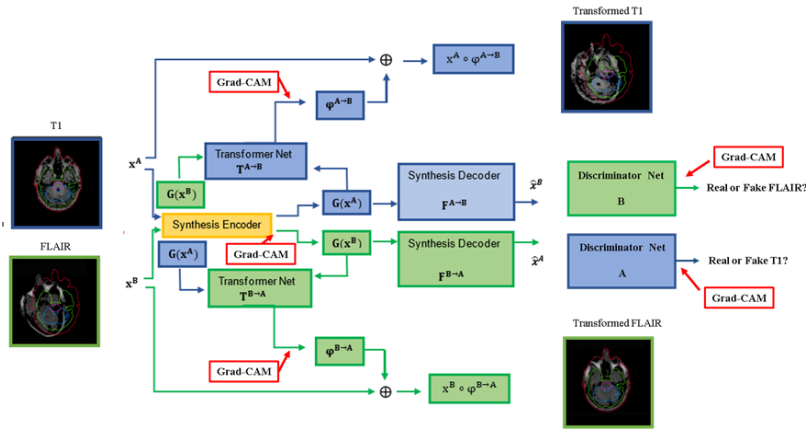

Figure 1) ExFIRE model architecture [16]. FIRE: Two synthesis encoders $G(x^A)$ and $G(x^B)$ extract modality-invariant (latent) representations. Two transformation networks $T^{A \to B}$ and $T^{B \to A}$ model the transformation fields. Two synthesis decoders $F^{A \to B}$ and $F^{B \to A}$ map the representations extracted by $G(x^A)$ and $G(x^B)$ to $\hat{x}^B$ and $\hat{x}^A$.

ExFIRE: Grad-CAM methodologies are incorporated within each of the model components to derive visual explanations regarding how each component contributes to unsupervised image registration. We generate attention maps for the last convolutional layers of: 1) the encoder network; 2) the spatial transformation networks; and 3) the discriminator networks. Specifically, the Grad-CAM attention maps of the encoder and transformation networks correlates the predicted deformation fields to the reference imaging space. The attention maps of discriminator networks can display the critical visual clues of the quality of synthesized images, thus further explain the performance of registration.

## 4. RESULTS AND DISCUSSION

We have previously demonstrated the robustness of our model on 3D and 2D brain (MRBrains13: http://mrbrains13.isi.uu.nl/; MRBrains18: https://mrbrains18.isi.uu.nl/) and cardiac 4D cardiac cine MRI data from the 2017 ACDC (https://www.creatis.insa-lyon.fr/Challenge/acdc) data [16]. In this work, we use 3D brain data to derive explainability maps across model components.

The dataset consisted of 3D $T_1$-weighted and $T_2$-Fluid-attenuated IR ($T_2$-FLAIR) data from 12 subjects, acquired using 3T MRI [16]. The 3D $T_1$ and $T_2$-FLAIR datasets included 192 and 48 slices per patient, respectively. The 3D $T_1$ and $T_2$-FLAIR datasets were already co-registered. Across all modalitites, images had a voxel size of $0.958 \times 0.958 \times 3.000$ mm$^3$. To assess model performance, manual annotations from three brain anatomical structures were used: the brain stem (BS), the cerebellum (Ce) and the white matter (WHM). For the training, validation and testing processes, we used all MRI slices from 8, 1 and 3 patients, respectively. To perform 3D registration, all data were resampled to 1.28mm$^3$ per voxel.

In the registration process, 3D registration was performed between $T_1$ and $T_2$-FLAIR data. During training, moderate-to-strong (20-50% change across at least one dimension) affine and non-rigid transformations were randomly applied to the moving and fixed images [16]. In the testing phase, the $T_1$ data were randomly transformed 20 times, and were subsequently allowed to be registered to the corresponding $T_2$-FLAIR data.

**Encoder Explainability**

Initially, we derive Grad-CAM visualizations in the encoder, which is designed to extract modality-invariant latent representations ($G(x^A)$, $G(x^B)$). Figure 2 shows the hybrid transformation fields (affine+non-rigid) that contribute more to the registration process, extracted from the encoder. Although the encoder learns modality-invariant latent representations, it is already obvious that the ExFIRE model gains complementary information from either imaging dataset ($T_1$ and $T_2$-FLAIR). Note that this complementary information can also be explained by the genuine design of these MRI modalities: $T_1$ is designed to provide rich anatomical information, whilst $T_2$-FLAIR suppresses anatomical information in an attempt to increase lesion to cerebrospinal fluid contrast (for lesion detection) [16, 25].

**Spatial Transformation Networks Explainability**

Subsequently, we derive Grad-CAM visualizations in the STN networks. Figure 3 shows the hybrid transformation fields that contribute more to the registration process. It is important to note again, that the ExFIRE model learns complementary information from either imaging dataset.

The observation that the ExFIRE models learns complementary information from either imaging modality in both the Encoder and STN components, means that the multi-modal transformations are invertable in relation to each other, which in turn proves that our method is intrinsically producing inverse-consistent topology-preserving image registrations.

**Discriminator Explainability**

Until now, we showed that the ExFIRE model learns and propagates complementary information from either imaging modality. Through this process, the proposed method learns and maintains inverse-consistency in the encoders and STN.

Finally, we derive Grad-CAM visualizations in the two discriminator networks (where the difference between the real and synthesized images are shown). Figure 4 illustrates that the ExFIRE model leverages homogeneous information from both the real and synthetic images: it is clear that Grad-CAM is either diffused (mainly in the $T_1$ analysis), or focusing on similar brain areas (mainly in the FLAIR analysis), across real and synthetic rigid and non-rigid registrations. The fact that the ExFIRE model extracts homogeneous information across real and synthetic images, reflects the quality of our image synthesis process, which in turn helps to optimize the cross-modal bi-directional registration within our network.

We should emphasise that $T_1$ contains higher anatomical information compared to FLAIR, and therefore benefits from leveraging diffuse (more homogeneous) information from both real and synthetic images, against FLAIR.

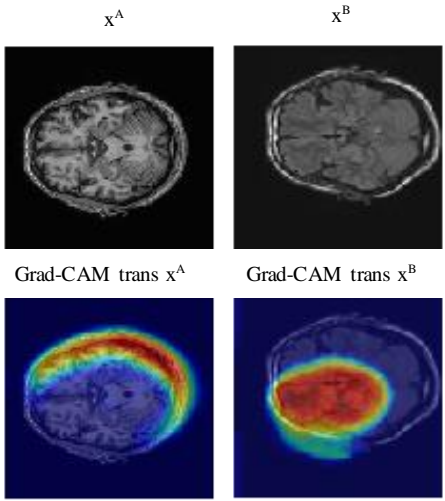

Figure 2) Grad-CAM results of hybrid transformation fields in the encoder, for $T_1$ ($x^A$) and FLAIR ($x^B$).

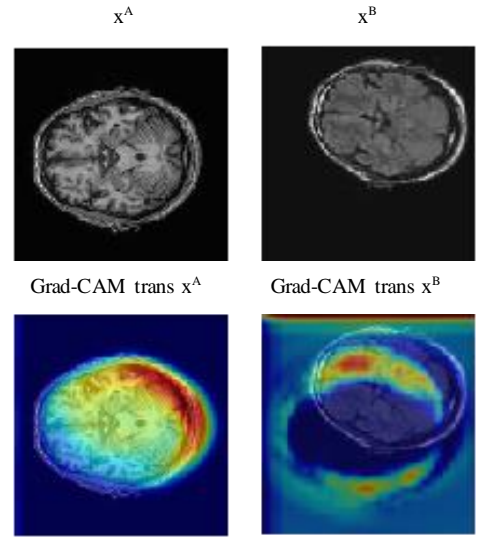

Figure 3) Grad-CAM results of hybrid transformation fields in the STN networks, for $T_1$ ($x^A$) and FLAIR ($x^B$). Note that transformation fields are centered for $x^B$.

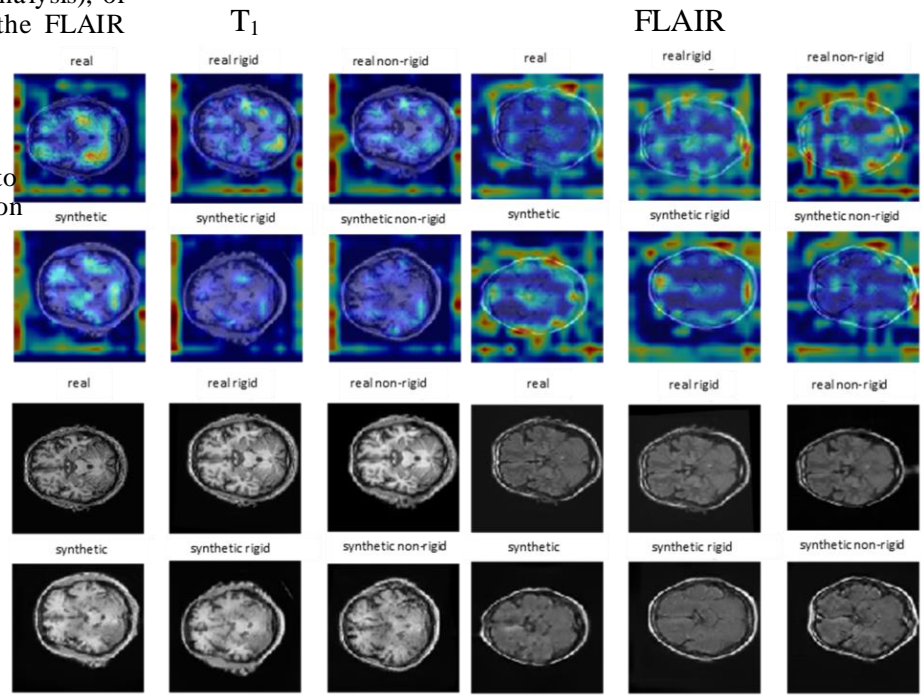

Figure 4) Grad-CAM results of rigid and non-rigid transformations from real and synthetic $T_1$ and FLAIR images.

**5. CONCLUSIONS**

We demonstrated that the ExFIRE learns complementary information from either imaging modality in the Encoder and STN components, showing that we produce inverse-consistent topology-preserving registrations. We also show that the ExFIRE model leverages homogeneous information from real and synthetic images in the decoders, during rigid and non-rigid transformations. We therefore demonstrate that our model is fully explainable, setting the framework to generalise our approach in the medical imaging domain.

# 6. REFERENCES

1. Rueckert D, Schnabel JA: Medical image registration. In: Biomedical image processing. Springer. 2010; 131–154.

2. Zhou SK, Greenspan H, Davatzikos C, et al. A review of deep learning in medical imaging: imaging traits, technology trends, case studies with progress highlights and future promises. Proceedings of the IEEE. 2021; 109(5): 820-838.

3. Fu Y, Lei Y, Wang T, Curran WJ, Liu T, Yang X. Deep learning in medical image registration: A review. Phys Med Biol. 2020; 65(20): 20TR01.

4. Haskins G, Kruger U, Yan P. Deep learning in medical image registration: A survey. Mach Vis Appl. 2020; 31: 8.

5. Cao X, Yang J, Zhang J, Nie D, Kim M, Wang Q, Shen D: Deformable image registration based on similarity-steered cnn regression. In: International Conference on Medical Image Computing and Computer-Assisted Intervention, Springer. 2017: 300-308.

6. Fan JF, Cao XH, Wang Q, Yap PT, Shen DG. Adversarial learning for mono- or multi-modal registration. Med. Image Anal. 58: 101545.

7. Fan JF, Cao XH, Yap EA, Shen DG. BIRNet: brain image registration using dual-supervised fully convolutional networks Med. Image Anal. 54: 193–206.

8. Elmahdy MS, et al. Robust contour propagation using DL and image registration for online adaptive proton therapy of prostate cancer. Med. Phys. 2019; 46: 3329-43.

9. Lei Y, Fu Y, Wang T, Liu Y, Patel P, Curran WJ, Liu T, Yang X. 4D-CT deformable image registration using multiscale unsupervised DL. 2020; Phys. Med. Biol. 65: 085003.

10. Vos B, Berendsen F F, Viergever M A, Staring M and Išgum I 2017 End-to-end unsupervised deformable image registration with a convolutional neural network. Deep Learning in Medical Image Analysis and Multimodal Learning for Clinical Decision Support. DLMIA 2017, ML-CDS. Lecture Notes in Computer Science, ed M Cardoso (Berlin: Springer). 2017; 10553. 204–12.

11. Ferrante E, Oktay O, Glocker B, Milone DH. On the adaptability of unsupervised CNN-based deformable image registration to unseen image domains. Machine Learning in Medical Imaging (MLMI). Lecture Notes in Computer Science vol 11046, ed Y Shi, H I Suk and M Liu (Berlin: Springer). 2018; 294–302.

12. de Vos BD, Berendsen FF, Viergever MA, Sokooti H, Staring M, Isgum I. A deep learning framework for unsupervised affine and deformable image registration. Medical image analysis. 2019; 52: 128-143.

13. Selvaraju RR, Cogswell M, Das A, Vedantam R, Parikh D, Batra D. GradCAM: visual explanations from deep networks via gradient-based localization. In: Proceedings of the IEEE International Conference on Computer Vision 2017, pp. 618–626.

14. Wei D, Ahmad S, Huo J, Huang P, Yap PT, Xue Z, Sun J, Li W, Shen D, Wang Q. SLIR: Synthesis, localization, inpainting, and registration for image-guided thermal ablation of liver tumors. Medical Image Analysis. 2020; 65.

15. Song X, Guo H, Xu X, Chao, et al. 2021. Cross-modal Attention for MRI and Ultrasound Volume Registration. MICCAI 2021, LNCS: pp. 66–75.

16. Wang C, Yang G, Papanastasiou G. FIRE: Unsupervised bi-directional inter- and intra-modality registration using deep networks. IEEE CBMS 2021.

17. Jaderberg M, Simonyan K, Zisserman A, et al. Spatial transformer networks. In: Advances in neural information processing systems. 2015; 2017-2025.

18. Balakrishnan G, Zhao A, Sabuncu MR, Guttag J, Dalca AV. Voxelmorph: a learning framework for deformable medical image registration. IEEE Trans Med Imaging. 2019; 38(8): 1788-1800.

19. Dalca AV, Balakrishnan G, Guttag J, Sabuncu MR. Unsupervised learning for fast probabilistic diffeomorphic registration. International Conference on Medical Image Computing and Computer-Assisted Intervention (MICCAI), Springer. 2018; 729-738.

20. Kim B, Kim DH, Park SH, Kim J, Lee JG, Ye JC. CycleMorph: Cycle consistent unsupervised deformable image registration. Med Image Anal. 2021; 71: 102036.

21. Avants BB, Epstein CL, Grossman M, Gee JC. Symmetric diffeomorphic image registration with cross-correlation: evaluating automated labeling of elderly and neurodegenerative brain. Medical image analysis. 2008; 12(1): 26-41.

22. Yang X, Kwitt R, Styner M, Niethammer M. Quicksilver: Fast predictive image registration-a deep learning approach. NeuroImage. 2017; 158: 378-396.

23. Nielsen RK, Darkner S, Feragen A. Topaware: Topology-aware registration. In: International Conference on Medical Image Computing and Computer-Assisted Intervention (MICCAI), Springer. 2019; 364–372.

24. Yang G, Ye Q, Xia J. Unbox the black-box for the medical explainable AI via multi-modal and multi-centre data fusion: A mini-review, two showcases and beyond. Information Fusion 2022; 77: 29-52.

25. Kates R, Atkinson D, Brant-Zawadzki M. Fluid-attenuated inversion recovery (FLAIR): clinical prospectus of current and future applications. Top Magn Reson Imaging. 1996;8(6):389-96.